\documentclass{article}
\usepackage{amsmath, amssymb,graphicx}

\begin{document}

\title{Measurement of the gluon polarization $\Delta G/G$ at COMPASS}

\author{C.~Schill (on behalf of the COMPASS collaboration)\\[1cm]
Albert-Ludwigs-Universit\"at Freiburg\\
Hermann-Herder-Str. 3, D-79104 Freiburg, Germany \\ 
E-mail: Christian.Schill@cern.ch}

\date{November 26, 2004}

\maketitle

\abstract{One of the key objectives of the COMPASS experiment at CERN is the
determination of the gluon contribution to the nucleon spin. The gluon
polarization is measured via photon-gluon fusion in deep-inelastic scattering
of $160$~GeV/c polarized muons on a polarized $^6$LiD solid-state target.
Photon-gluon fusion  is tagged by the observation of charmed mesons or the
production of hadron pairs with large transverse momenta $p_t$. The status of
the analysis of the $D^0$ and $D^{0*}$ events and of the high-$p_t$
hadron pairs is shown. The gluon polarization $\Delta G/G$ has been determined
from the asymmetry of high-$p_t$ hadron pairs with $Q^2>1$~(GeV/c)$^2$ in an
analysis of the 2002/03 data as  $\Delta G/G=0.06 \pm 0.31 (stat.) \pm 0.06
(syst.)$.}

\section{Introduction}

In an intuitive picture the spin of the nucleon is carried by its valence
quarks. However, deep inelastic scattering (DIS) experiments  (EMC, SMC, SLAC,
HERMES) have shown that only a small fraction of the nucleon spin is carried by
quarks. Since then, it has been one of the key questions in hadron physics, how
the total spin of the nucleon $\hbar/2$ is composed. Candidates, which may
contribute in addition to the quark spin $\Delta \Sigma$ are the helicity
contribution of the gluon $\Delta G$ as well as the quark and
gluon orbital angular momenta $L_{q,g}$:
\begin{equation}
\frac{1}{2}=\frac{1}{2}\Delta \Sigma + \Delta G + L_{q,g}
\end{equation}
One of the main goals of the COMPASS experiment\cite{co} is a measurement of
the helicity contribution $\Delta G$ of the gluon to the nucleon spin, via the
photon-gluon fusion process. Photon-gluon fusion is tagged by the 
production of charmed mesons $D^0$ and $D^{0*}$ or of hadron
pairs with large transverse momenta $p_t$. In addition to the gluon
polarization, the COMPASS experiment investigates a broad physics
program\cite{compass} in polarized semi-inclusive deep inelastic scattering and
hadron spectroscopy.

\section{The COMPASS experiment}

The COMPASS experiment uses a $160$~GeV/c polarized muon beam of the CERN SPS
scattering it off a polarized $^6$LiD solid state target\cite{target} at a
high luminosity of about $4\cdot 10 ^{32}$~cm$^{-2}$s$^{-1}$. From the
counting rate difference in two oppositely polarized target cells, the
photon-nucleon 
cross-section asymmetry $A^{\gamma^*d}$ can be determined:
\begin{equation}
A^{\gamma^*d}=\frac{\Delta\sigma^{\gamma^* d}}{\sigma^{\gamma^* d}}=
\frac{1}{P_bP_tfD}\cdot
            \frac{N_1^\leftrightarrows-N_2^\leftleftarrows}
	    {N_1^\leftrightarrows+N_2^\leftleftarrows},
\end{equation}
where the muon beam polarization is $P_b\approx 0.76$, the target polarization 
$P_t\approx 0.5$, and the fraction of polarized material in the target $f\approx
0.4$. The depolarization factor $D(y)$ of the virtual photon $\gamma^*$ can be calculated
as a function of the fractional energy transfer $y$.

The particles produced in the interaction are detected in a two-stage forward
spectrometer with high momentum resolution, high rate capability and an
excellent particle identification using hadronic and electromagnetic
calorimeters and a large ring-imaging \v{C}erenkov detector, which is able to
identify kaons and pions from the charmed meson decay. A special quasi-real
photo-production trigger allows to detect events with scattered muons down to
$Q^2 =10^ {-4}$~(GeV/c)$^2$. Data have been taken from 2002 until 2004 so far.

\section{Gluon polarization from open charm production}

The ``golden channel'' to tag photon-gluon fusion events is the production of a
$c$$\overline{c}$-quark pair, since the  charm content of the nucleon is very
small and the hard scale is set by the charm mass for this channel. One of the
charm quarks fragments into a  $D^0$ or a $D^{0*}$ meson, which is detected in
our experiment. The  $D^0$ and $D^{0*}$ mesons are reconstructed from their
invariant mass in the decays $D^0 \rightarrow K^- \pi^+$ and $D^{0*}\rightarrow
D^0 + \pi$ and the charge-conjugated decays (Fig.~\ref{figure}). The gluon
polarization $\Delta G/G$ can then be determined from the experimental
asymmetry in the open-charm production according to:
\begin{equation}
A^{\gamma^*d\rightarrow c \overline{c}}=\frac{\int{d \hat{s} \Delta \sigma^{PGF}
(\hat{s})\Delta G(x_g,\hat{s})}}
{\int{d \hat{s} \sigma^{PGF}(\hat{s}) G(x_g, \hat{s})}}\approx\left<a_{LL}\right>
\frac{\Delta G}{G},
\end{equation}
where $\hat{s}$ is the invariant mass $m^2_{c\overline{c}}$ of the charm quark
pair. The polarized photon-gluon cross section
$\Delta \sigma^{PGF}$ has been calculated in NLO by two
groups\cite{bojak,grispos}. The analysis of the full 2002-2004 dataset
is in progress, the projected statistical error on $\Delta G/G$ is 0.24.

\section{Gluon polarization from hadron pairs with large $p_t$}

\begin{figure}[t]
\includegraphics[width=5.8cm,height=4.9cm]{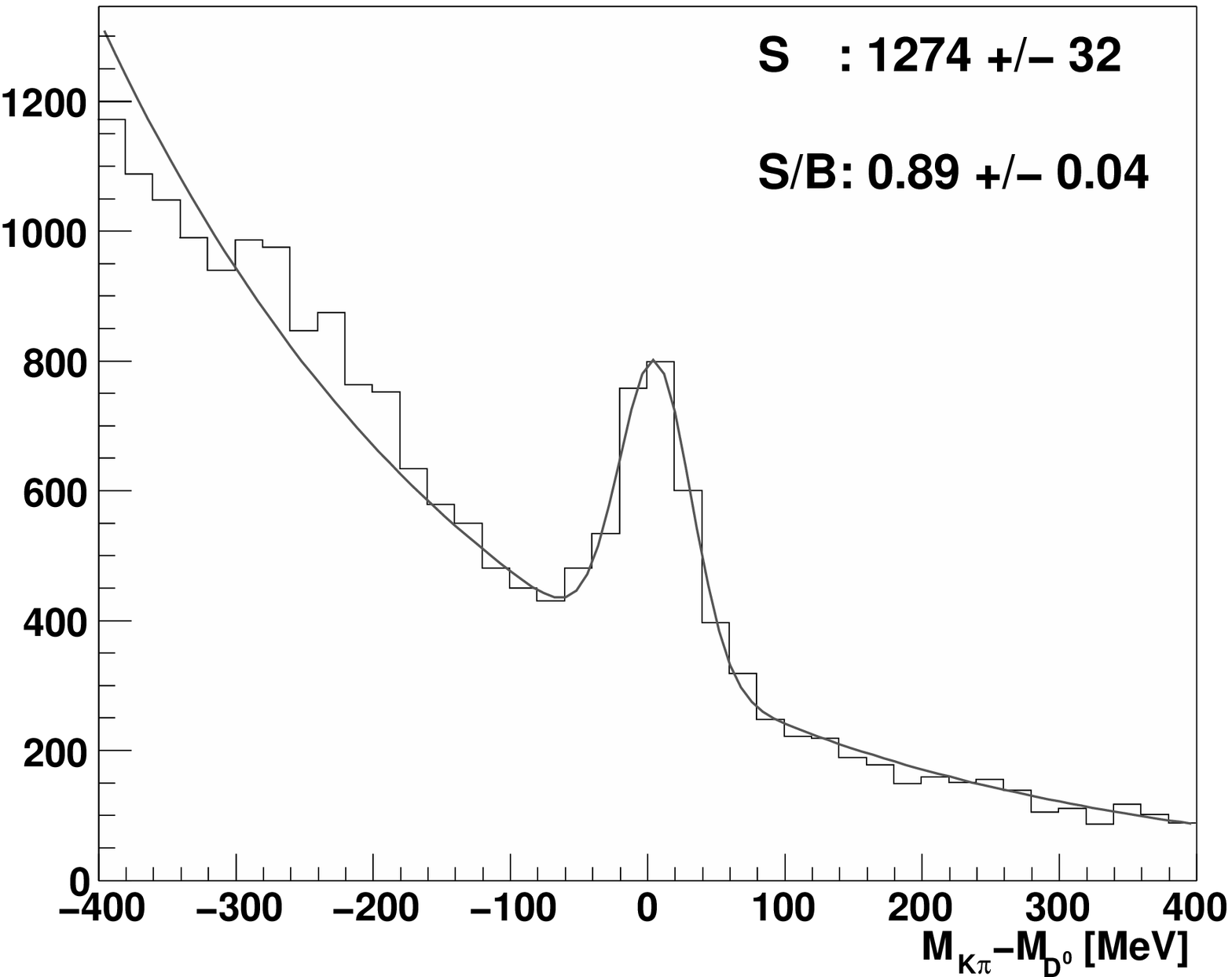}\\[-5.5cm]
\hspace*{5.8cm}
\includegraphics[width=5.8cm,height=5.84cm]{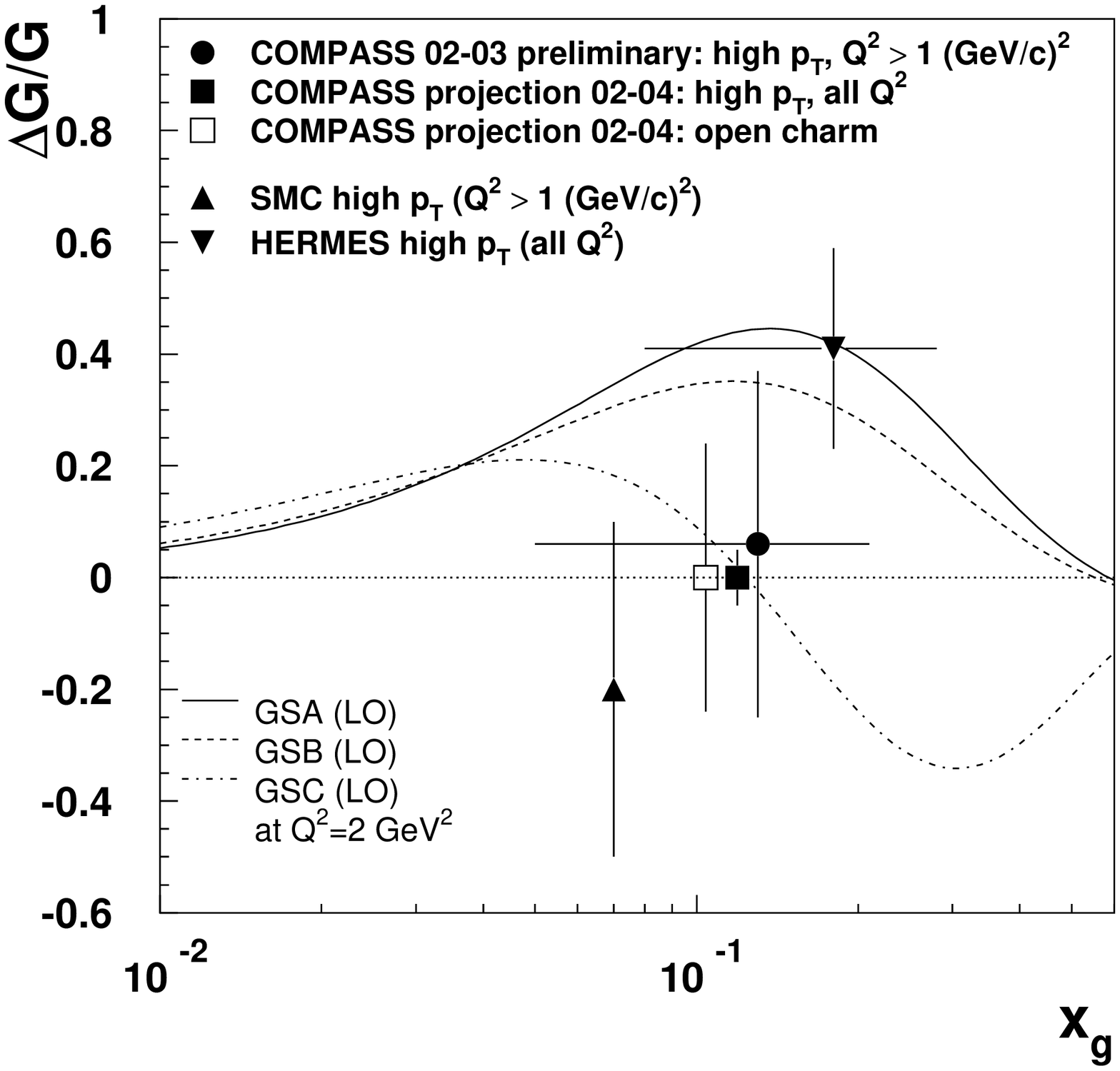}\\[-8.6mm]
\caption[]{Left: Reconstructed $D^0$ in the $K\pi$ invariant mass spectrum 
tagged by their decay from $D^*$ (2003 data). Right: COMPASS result for 
$\Delta G/G$ from high-$p_t$ hadron pairs with $Q^2>1$~(GeV/c)$^2$ from 2002/03 data 
compared to SMC\cite{smc} ($Q^2>1~(GeV/c)^2$) and HERMES\cite{avetik} (all Q$^2$)  measurements. In
addition, projections for the statistical accuracy in $\Delta G/G$ from open
charm production and high-$p_t$ hadron pairs for all $Q^2$ are shown for  the
full 2002-04 dataset. The curves show the parameterizations A - C of \cite{sterling}.}
\label{figure}
\end{figure}

Another approach to tag the photon-gluon fusion process is the detection of a
hadron pair\cite{bravar} with large transverse momenta $p_t$. The transverse
momentum of each hadron relative to the virtual photon is required to be larger than $0.7$~GeV/c and
$(p_{t1}^2+p_{t2}^2 )>2.5$~(GeV/c)$^2$. To ensure that the hadrons originate
from the current fragmentation region cuts on $x_F>0.1$ and $z>0.1$ have been
applied. Contributions from resonances are removed by a two hadron invariant
mass cut $m(h_1h_2)>1.5$~GeV/c$^2$. Requiring $Q^2>1$~(GeV/c)$^2$ suppresses a
possible contribution from resolved photon processes,  where the hadronic
structure of the photon is probed, and requiring $x_{Bj}<0.05$ selects a
kinematic region, where the asymmetry from leading order DIS and QCD-Compton
scattering is small.  From the selected high-$p_t$ sample of events in the
2002/2003 data we have measured: 
\begin{equation} 
A_{LL}^{\gamma^*d\rightarrow hhX}=-0.015 \pm 0.080(stat.) \pm 0.013 (syst.)
\end{equation}
The systematic uncertainty takes into account possible false asymmetries, the
uncertainty in the measurement of the target and beam polarization and the
knowledge of the depolarization $D$ and dilution factor $f$. The gluon
polarization is calculated from the asymmetry $ A_{LL}^{\gamma^*d\rightarrow
hhX}$ as:
\begin{equation}
A_{LL}^{\gamma^*d\rightarrow hhX}=\left<\frac{\hat{a}_{LL}^{PGF}}{D}\right>
\frac{\sigma^{PGF}}{\sigma^{tot}}\frac{\Delta G}{G},
\end{equation}
where $\hat{a}_{LL}^{PGF}$ is the analyzing power and $\sigma^{PGF}/\sigma^{tot}$ the 
 fraction of photon-gluon fusion events. Background processes like
QCD-Compton scattering and leading order DIS contribute only as a dilution to
the measured signal:  their asymmetry is proportional to $A_1^d(x)$, which is very
small in the selected kinematic range $x_{Bj}<0.05$. Their effect has been taken into
account in the systematic uncertainty of the result.
The analyzing power $\hat{a}_{LL}^{PGF}=-0.75\pm0.05(syst.)$ and the 
fraction $\sigma^{PGF}/\sigma^{tot}=0.34\pm0.07(syst.)$ of photon-gluon fusion
events were determined using a Monte-Carlo simulation (LEPTO) with a modified
set of fragmentation parameters\cite{smc} and including radiative corrections
(RADGEN). Our result for $\Delta G/G$ is (Fig.~\ref{figure}):
\begin{equation}
\Delta G/G=0.06 \pm 0.31 (stat.) \pm 0.06 (syst.)
\end{equation}
at a mean gluon momentum fraction $\left<x_g\right>=0.13\pm0.08$(RMS).

\section{Outlook}

The first COMPASS result on $\Delta G/G$ for high-$p_t$ hadron pairs with
$Q^2>1$~(GeV/c)$^2$ is shown in Fig.\ref{figure} in comparison with other
experiments\cite{smc,avetik}. Including the 2004 run, the present data will be
approximately doubled. There are about 10 times more events at
$Q^2<1$~(GeV/c)$^2$. However, at low $Q^2$, a background of resolved photons
enters as an additional theoretical uncertainty. The projected statistical
accuracy on $\Delta G/G$ from open charm production and high-$p_t$ hadron pairs
for all $Q^2$ is shown in Fig.\ref{figure}. COMPASS will resume data taking
in 2006 and will continue its physics program until at least 2010.

\end{document}